\documentclass[%
 reprint,
 amsmath,amssymb,
 aps,
]{revtex4-2}

\usepackage{lineno}
\usepackage{xcolor}
\usepackage{soul}
\usepackage{subfigure}
\usepackage{float}
\usepackage{lineno}
\usepackage{tabularx}
\usepackage{graphicx}
\usepackage{dcolumn}
\usepackage{bm}

\begin{document}

\preprint{APS/123-QED}

\title{Analysis and design of two-dimensional compound metallic metagratings using an analytical method}
\author{Mahdi Rahmanzadeh}

\author{Amin Khavasi}%
 \email{khavasi@sharif.edu}
\affiliation{Electrical Engineering Department, Sharif University of Technology, Tehran 11155-4363, Iran}


\begin{abstract}
The recently proposed concept of metagrating enables wavefront manipulation of electromagnetic (EM) waves with unitary efficiency and relatively simple fabrication requirements. Herein, two-dimensional (2D) metagratings composed of a 2D periodic array of rectangular holes in a metallic medium are proposed for diffraction pattern control. We first present an analytical method for diffraction analysis of 2D compound metallic metagrating (a periodic metallic structure with more than one rectangular hole in each period). Closed-form and analytical expressions are presented for the reflection coefficients of diffracted orders for the first time. Next, we verify the proposed method's results against full-wave simulations and demonstrate their excellent agreement. As a proof of principle, two applications are presented using the proposed analytical method. The first application is a perfect out-of-plane reflector that transfers a normal transverse-magnetic (TM) polarized plane wave to an oblique transverse-electric (TE) polarized plane wave in the $y-z$ plane. The second one is a five-channel beam splitter with an arbitrary power distribution between channels. Using the proposed analytical method, we designed these metagratings without requiring even a single optimization in a full-wave solver. The performance of the designed metagratings is better than previously reported structures in terms of power efficiency and relative distribution error. Our analytical results reveals that 2D metagratings can be used for manipulating EM waves in the plane and out of the plane of incidence with very high efficiency, thereby leading to extensive applications in a wide range of frequencies from microwave to terahertz (THz) regimes.
\end{abstract}

\maketitle


\section{Introduction}

Wavefront shape manipulation has always been an interesting topic in electromagnetism due to its fundamental role in several applications such as radars, imaging, and communication systems \cite{novotny2012principles}. In the last two decades, artificial structures have reached the peak of attention for controlling electromagnetic waves\cite{pendry2000negative,pendry2006controlling,engheta2006metamaterials,joannopoulos1997photonic,yang2019surface}. Metasurfaces, in particular, have been used to devise novel devices with significant practical and scientific applications thanks to their high potential\cite{kildishev2013planar,epstein2016huygens,rahmanzadeh2018multilayer,hosseininejad2019reprogrammable,hadad2015space,mueller2017metasurface,momeni2019generalized,rajabalipanah2020real}. Metasurfaces are two-dimensional thin planar patterned structures formed by spatially arranged building blocks called meta-atoms. They can be designed to control the amplitude, phase, and polarization of EM waves. A major class of metasurfaces is gradient metasurfaces (GMS) which can manipulate EM waves by imparting local momentum to the incoming EM waves through the gradient phase profile of the structure\cite{yu2011light}. GMS can realize a wide range of electromagnetic and optical functionalities from beam focusing to holographic imaging \cite{aieta2012aberration,rouhi2019multi,yu2014flat,sun2012gradient,ni2013metasurface}. However, passive and local GMS have been shown to suffer from low power efficiency and require precise and high-resolution fabrication processes\cite{asadchy2016perfect,estakhri2016wave,diaz2017generalized}. These problems restrict many of the applications mentioned above. To address these problems, the concept of metagrating was proposed by Ra'di \textit{et al} \cite{ra2017metagratings}.

Metagratings, a sparse periodic array of subwavelength scatterers (meta-atoms), have attracted considerable interest in the last few years because they allow the realization of diverse phenomena such as anomalous reflection, beam splitting, beam steering, and beam focusing\cite{ra2018reconfigurable,rahmanzadeh2021analyticaloe,xu2021analysis,popov2019designing,rahmanzadeh2021analytical,rabinovich2018analytical,rajabalipanah2021analytical,kang2020efficient,epstein2017unveiling}. Their power efficiency is not restricted by any fundamental physical bounds, and they require much less fabrication complexity than metasurfaces\cite{ra2017metagratings}. The working principle of metagratings can be understood using Floquet-Bloch (FB) theory, according to which when a plane wave impinges to a periodic structure, it will be diffracted into several discrete waves in a certain direction. Meta-atom properties have a significant effect on diffracted waves (FB mode); hence, by engineering meta-atoms, we can tailor the desired diffraction patterns. Different geometries for meta-atoms, such as loaded thin wires\cite{popov2019constructing}, one-dimensional (1D) grooves\cite{rahmanzadeh2020perfect}, and graphene ribbons \cite{behroozinia2020real} have been used to design metagratings to realize various functionalities.

Most of the designed metagratings are periodic in 1D and are sensitive to incident wave polarization. Therefore 2D-metagratings (periodic array in two directions) are proposed for realizing polarization-independent anomalous reflection with high diffraction efficiency \cite{chen2018polarization,rabinovich2020dual,zhou2020polarization}. However, 2D metagratings have not been designed based on an analytical method and thus have a time-consuming design procedure. For example, in \cite{rabinovich2020dual}, an all-metallic metagrating is proposed for polarization-independent perfect anomalous reflection. It is analyzed by the rigorous coupled-wave analysis (RCWA) technique and anomalously reroutes an obliquely incident wave with high efficiency for both polarization. However, the RCWA technique and other numerical methods can not present a closed-form expression for the reflection coefficients; therefore, an analytical method for the analysis of 2D metagratings is in demand for accelerating design procedures. Furthermore, all of the existing metagratings are used for in-plane control of EM waves (The wave-vector of the diffracted wave lies in the plane of incidence). Although \cite{inampudi2018neural} has attempted to transfer the incident power to some directions out of the plane of the incidence, due to its sophisticated method design, the designed metagratings have a low power efficiency. Out-of-plane manipulation of EM waves (The wave-vector of the diffracted wave does not lie in the plane of incidence) has interesting potential applications such as 2D planar lenses with high numerical apertures, flat polarization converters, and radar cross-section reductions\cite{aieta2012out,aieta2012reflection,rajabalipanah2018circular,liu2018negative}. Metasurfaces are common elements for realizing this phenomenon, which, as noted earlier, have low power efficiency. To the best of the authors' knowledge, out-of-plane manipulation of EM waves has not been realized by metagratings. 

In this paper, we present an analytical method for analyzing two-dimensional compound metallic metagratings (2D-CMGs) and show these metagratings enable in-plane and out-of-plane EM wave control. The proposed metagrating consists of the 2D-periodic repetition of a finite number of rectangular holes carved out of a thick metal slab. To derive our method, we first expand the electromagnetic field by FB theory and extract the reflection coefficients of the zeroth diffracted order and higher diffracted orders by using appropriate boundary conditions in the conjunction mode-matching technique. The accuracy of the proposed method is verified through numerical examples. Using this analytical method, we designed an out-of-plane anomalous reflector with unitary efficiency at normal incidence. Next, we propose a five-channel beam splitter using 2D-CMG. Two of these channels are in the plane perpendicular to the plane of incidence. The power distribution between these channels are arbitrary. The performance of the designed beam splitters was better than that of the previously reported metasurface- and metagratings-based beam splitters. Finally, some practical aspects of the experimental realization of the designed devices are discussed.

\section{Analytical method for analysis of 2D-CMGs}

\begin{figure} 
\centering\includegraphics[width=\linewidth]{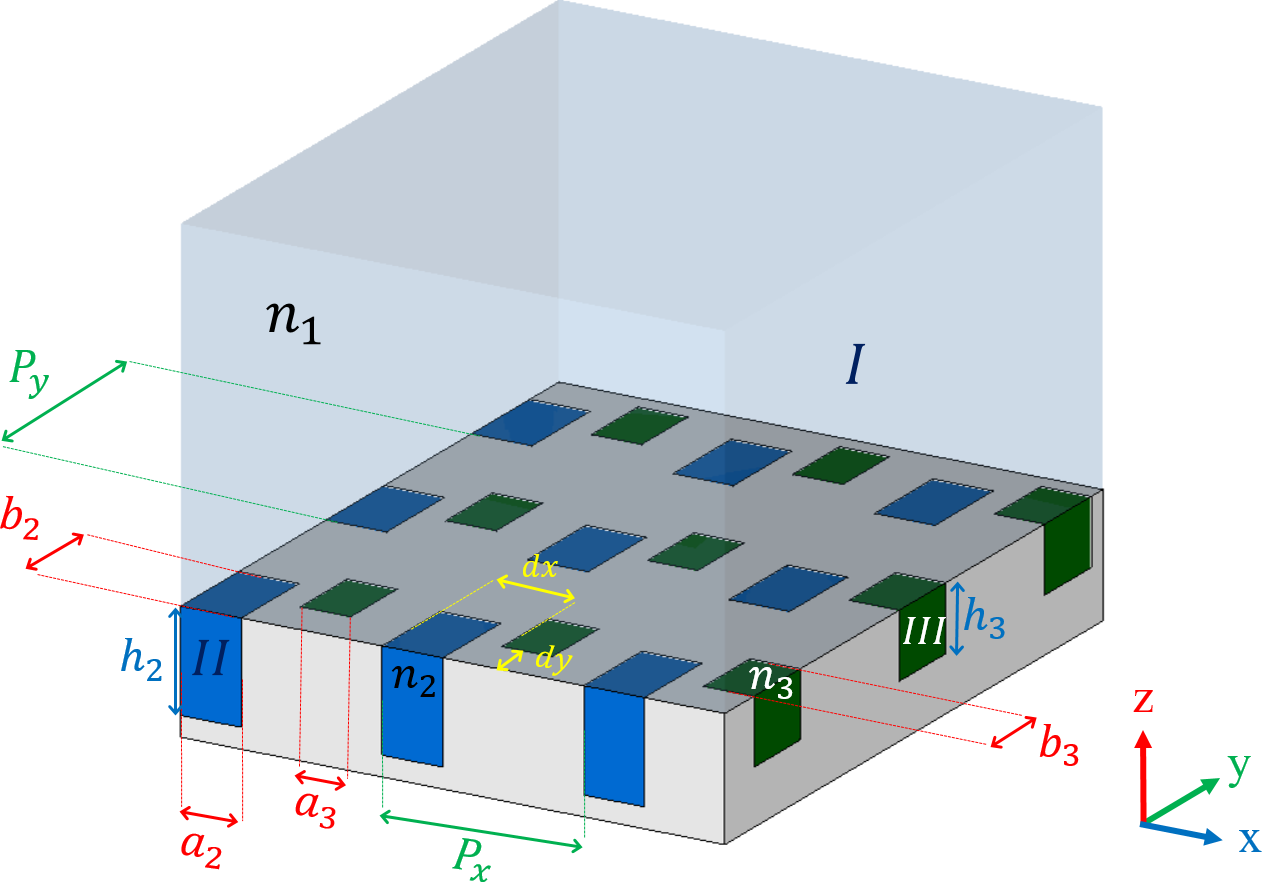}
\caption{The structure of CMG with two rectangular holes per period. CMG is covered by a homogeneous medium.}
\label{CMG_twoH}
\end{figure}

\subsection{Diffraction of a normal TM polarized plane wave by 2D-compound metallic grating}

Herein, we first present an analytical and closed-form expression for the reflection coefficients of diffracted orders of 2D-CMG comprising two rectangular holes. Next, we discuss how the proposed method can be generalized to analyze a 2D-CMG containing an arbitrary number of rectangular holes. It should also be noted that the time-harmonic of the form $\exp(j{\omega}t)$ is assumed throughout this paper.

 Consider a 2D-CMG including two rectangular holes made in a metallic medium as shown in Fig.~\ref{CMG_twoH}. The periods of the structure along the $x$- and $y$-axes are $P_x$ and $P_y$, respectively. Each hole has a width of $a_i$, length of $b_i$, and height of $h_i$, and is filled with a dielectric medium with a refractive index of $n_i$(region II and III). The whole structure is surrounded by a homogeneous medium with a refractive index of $n_1$ (region I). We denote the distance between the corner of holes with $d_x$ and $d_y$ in the $x$- and $y$-directions, respectively.

Assume that a normal incident TM polarized plane wave (the magnetic field in the $y$-direction) propagating along the $z$-direction illuminates the structure. The tangential electric and magnetic fields in the region $z>0$ can be written as \cite{marques2009analytical,khavasi2015corrections}

\begin{widetext}
\begin{eqnarray} \label{Ex_1}
E_{1x} = e^{j{k_{z,00}}z} +R_{00}{e^{ - j{k_{z,00}}z}}  +  \sum\limits_{m \ne 0} R_{m0}^{TM}e^{- j{k_{x,m}}x} e^{ - j{k_{z,m0}}z}+  \sum\limits_{n \ne 0} R_{0n}^{TE}\ e^{ - j{k_{y,n}}y}e^{ - j{k_{z,0n}}z} \nonumber \\ +\sum\limits_{m\ne0} \sum\limits_{n \ne 0} (R_{mn}^{TM} +  R_{mn}^{TE})e^{-j(k_{x,m}x+k_{y,n}y+k_{z,mn}z)} 
\end{eqnarray}

\begin{eqnarray} \label{Hy_1}
H_{1y} =  - \xi _{00}e^{j{k_{z,00}}z} + \xi _{00} R_{00} e^{-j{k_{z,00}}z}+\sum\limits_{m \ne 0} \xi _{m0}R_{m0}^{TM}e^{ - j{k_{x,m}}x} e^{- j{k_{z,m0}}z}+ \sum\limits_{n \ne 0} \xi _{0n}^{TE}R_{0n}^{TE}\ e^{ - j{k_{y,n}}y} e^{-j{k_{z,0n}}z}  \nonumber \\+ \sum\limits_{m \ne 0} \sum\limits_{n \ne 0} (\xi _{mn}^{TM}R_{mn}^{TM}+  \xi _{mn}^{TE}R_{mn}^{TE}) e^{-j(k_{x,m}x+k_{y,n}y+k_{z,mn}z)}
\end{eqnarray}
\end{widetext}

Using Maxwell equations, the other components of the tangential fields are obtained as \cite{balanis2012advanced}

\begin{eqnarray} \label{Ey_1}
 E_{1y} = \sum\limits_{m \ne 0} \sum\limits_{n \ne 0} (-\frac{k_{x,m}}{k_{y,n}}R_{mn}^{TE} \nonumber\\ +\frac{k_{y,n}}{k_{x,m}}R_{mn}^{TM}) e^{-j(k_{x,m}x+k_{y,n}y+k_{z,mn}z)}
\end{eqnarray}

\begin{eqnarray} \label{Hx_1}
  H_{1x} =\sum\limits_{m \ne 0} \sum\limits_{n \ne 0} (\frac{k_{x,m}}{k_{y,n}}\xi _{mn}^{TE}R_{mn}^{TE} \nonumber\\ -\frac{k_{y,n}}{k_{x,m}}\xi _{mn}^{TM} R_{mn}^{TM}) e^{-j(k_{x,m}x+k_{y,n}y+k_{z,mn}z)} 
\end{eqnarray}

where $R_{mn}^{TE}$ and $R_{mn}^{TE}$ are the reflection coefficients of the TE- and TM-polarized $mn$th diffracted order, respectively and the subscripts $m$ , and $n$ correspond to the order of the diffracted waves along the $x$- and $y$-axes, respectively. Furthermore, $k_{x,m}$, $k_{y,n}$, and $k_{z,mn}$ are the wave-vector components of the diffraction order along the $x$- , $y$-, and $z$-directions in region I, respectively, and are given by \cite{khavasi2015corrections}

\begin{subequations}
\label{kxm1_kyn1-kz1}

  \begin{equation}
  \label{kxm}
{k_{x,m}} = \frac{{2m\pi }}{{{P_x}}}\,\,\,;\,\,\,m = 0,\,\, \pm 1,\, \pm 2,\,\,...
  \end{equation}
 
   \begin{equation}
  \label{kyn}
{k_{y,n}} = \frac{{2n\pi }}{{{P_y}}}\,\,\,\,;\,\,\,\,n = 0,\,\, \pm 1,\, \pm 2,\,\,...
  \end{equation}
     \begin{equation}
  \label{kzmn}
{k_{z,mn}} =-j\sqrt {k_{x,m}^2 + k_{y,n}^2 - k_0^2n_1^2};~m,n =0,\pm1,\pm2,...
  \end{equation}
\end{subequations}
where $k_0 =\omega(\varepsilon_0 \mu_0)^{1/2}$ is the free space wavenumber. It should be noted that the branch of the square root for the $z$-component of the wave-vector is chosen in such a way that either its real part should be positive (propagating wave) or its imaginary part should be negative (evanescent wave). Moreover, $\xi_{mn}^{TM}=\omega \varepsilon_0 n_1^2 / k_{z,mn} $ and $\xi_{mn}^{TE}=k_{z,mn}/\omega \mu_{0} $ are the TM/TE-wave admittance of the $mn$th diffracted order in region I.

In regions II and III, we assume that the holes are single-mode, and due to the TM polarization of the incident wave, we only take into account the $TE_{01}$ mode, which is propagating inside the holes, while assuming that the effects of other order modes are negligible. The validity of this approximation is limited to the operating frequency less than $f_c=\min[\sqrt{(\pi/a_i)^2+(\pi/b_i)^2}, \sqrt{(2\pi/b_i)^2}]/\sqrt{4\pi^2\mu_0\varepsilon_0 n_i^2}$ $(i=2,3)$, where the higher modes inside the holes are evanescent. As a result, the magnetic and electric fields in the holes can be written as \cite{balanis2012advanced}

\begin{subequations} \label{EandH2}
\begin{eqnarray} \label{Ex2}
    {E_{2x}} = T_2^ - \,\sin (\frac{\pi }{{{b_2}}}y){e^{j{\beta _{2,1}}z}} \nonumber\\- T_2^ - \,{e^{ - 2j{\beta _2}{h_2}}} \,\sin (\frac{\pi }{{{b_2}}}y){e^{ - j{\beta _{2,1}}z}}
\end{eqnarray}

\begin{eqnarray} \label{Hy2}
    H_{2y} =  - T_2^ - \,\xi _2^{TE}\sin (\frac{\pi }{{{b_2}}}y){e^{j{\beta _2}z}} \nonumber\\ - T_2^ - \,{e^{ - 2j{\beta _2}{h_2}}} \xi _2^{TE}\,\sin (\frac{\pi }{{{b_2}}}y){e^{ - j{\beta _2}z}}
\end{eqnarray}
\end{subequations}

for $x\in[0, a_2]$,$y\in[0, b_2]$, and
\begin{subequations} \label{EandH3}
\begin{eqnarray} \label{Ex3}
    {E_{3x}} = T_3^ - \,\sin (\frac{{\pi (y - {d_y})}}{{{b_3}}}){e^{j{\beta _3}z}} \nonumber\\ - T_3^ - \,{e^{ - 2j{\beta _3}{h_3}}}\,\sin (\frac{{\pi (y - {d_y})}}{{{b_3}}}){e^{ - j{\beta _3}z}}
\end{eqnarray}
\begin{eqnarray} \label{Hy3}
    H_{3y} = -T_3^ - \,\xi_3^{TE}\sin(\frac{\pi (y - d_y)}{b_3})e^{j{\beta _3}z} \nonumber\\ - T_3^ - \,e^{-2j \beta _3 h_3} \xi_3^{TE}\,\sin(\frac{\pi (y - {d_y})}{b_3})e^{-j \beta _3 z}
\end{eqnarray}
\end{subequations}

for $x\in[d_x, d_x+a_3]$,$y\in[d_y, d_y+b_3]$ and $\beta_i=\sqrt{(n_ik_0)^2-(\pi/b_i)^2}$ $(i=2,3)$  is the propagation constant of the $TE_{01}$ mode supported by a rectangular waveguide. In addition, $\xi=\beta_i/\omega \mu_0$ $(i=2,3)$ is the wave admittance of the $TE_{01}$ mode inside each hole.

Now, applying the boundary conditions at $z=0$ for the electric fields (the continuity of $E_x$ and $E_y$ at every point of the unit cell) leads to the following equations

\begin{subequations} \label{Ref_CO-a}
\begin{equation} \label{R_00-1}
    1 + {R_{00}} = {f_2}{S_2}A_{00}^{ + ,2}T_2^ -  + {f_3}{S_3}A_{00}^{ + ,3}\,T_3^ -
\end{equation}
\begin{equation} \label{R_m0-1}
    R_{m0}^{TM} = {f_2}{S_2}A_{m0}^{ + ,2}T_2^ -  + {f_3}{S_3}A_{m0}^{ + ,3}\,T_3^ - \,\,\,\,m \ne 0
\end{equation}
\begin{equation} \label{R_0n-1}
    R_{0n}^{TE} = {f_2}{S_2}A_{0n}^{ + ,2}\,T_2^ -  + {f_3}{S_3}A_{0n}^{ + ,3}\,T_3^ - \,\,\,\,n \ne 0
\end{equation}

\begin{eqnarray} \label{R_mn-1}
    R_{mn}^{TE} = R_{mn}^{TM}\frac{{k_{y,n}^2}}{{k_{x,m}^2}} = \frac{k_{y,n}^2}{k_{x,m}^2 + k_{y,n}^2}( \nonumber\\ f_2 S_2 A_{mn}^{+,2}\,T_2^ -  + f_3 S_3 A_{mn}^{+,3}\,T_3^-)\,\,\,m,n \ne 0
\end{eqnarray}
\end{subequations}

 wherein $f_i=a_i b_i/P_x P_y$, $S_i=1-e^{-2j\beta_i h_i}$ $(i=2,3)$, and
 \begin{eqnarray} \label{Am}
A_{mn}^{\pm,i} =\frac{1}{a_i b_i} \int\limits_{d_{xi}}^{d_{xi} +a_i} \int\limits_{d_{yi}}^{d_{yi} + b_i} \sin (\frac{\pi (y-d_{yi})}{b_i})\nonumber\\  \times e^{\pm j(k_{x,m}x+k_{y,n}y)} dy dx ~~;~~~i = 2,3\nonumber\\;~{d_{x2}}=d_{y2}=0, d_{x3}=d_x, d_{y3} = d_y
\end{eqnarray}

which are obtained by multiplying the electric fields to ${e^{j{k_{x,m}}x}}\,{e^{j{k_{y,n}}y}}$ and taking the integral of over one unit cell. Similarly, we apply the continuity of the tangential magnetic fields ($H_x$ and $H_y$) at $z = 0$. Using \eqref{Ex_1},\eqref{Ex2} and \eqref{Ex3}, and by multiplying the magnetic fields by $\sin (\pi(y-d_{yi})/b_i)$ and then taking the integral of both sides over each hole, we have

   \begin{subequations} \label{BCforH}
   \begin{eqnarray} \label{BCforH-hole1}
- {\xi _{00}}A_{00}^{ - ,2} + {\xi _{00}}A_{00}^{ - ,2}{R_{00}} + \sum\limits_{m \ne 0} {\xi _{m0}^{TM}A_{m0}^{ - ,2}R_{m0}^{TM}}  \nonumber\\ + \sum\limits_{n\ne0} {\xi _{0n}^{TE}A_{0n}^{ - ,2}R_{0n}^{TE}}+ \sum\limits_{m \ne 0} \sum\limits_{n \ne 0} A_{mn}^{ - ,2}(\xi _{mn}^{TM}R_{mn}^{TM} \nonumber\\ + \xi _{mn}^{TE}R_{mn}^{TE})=-0.5 S'_2 \,T_2^-
   \end{eqnarray}
   
     \begin{eqnarray} \label{BCforH-hole2}
 -\xi_{00} A_{00}^{-,3} +\xi_{00} A_{00}^{- ,3} R_{00} + \sum\limits_{m \ne 0}\xi _{m0}^{TM}A_{m0}^{-,3}R_{m0}^{TM} \nonumber \\ + \sum\limits_{n \ne 0} \xi_{0n}^{TE}A_{0n}^{-,3}R_{0n}^{TE} + \sum\limits_{m\ne0}\sum\limits_{n \ne 0} A_{mn}^{-,3}(\xi _{mn}^{TM} R_{mn}^{TM} \nonumber \\ +\xi _{mn}^{TE} R_{mn}^{TE})= -0.5 S'_3 T_3^-
   \end{eqnarray}
   \end{subequations}
where $S'_i=\xi _i^{TE} (1 +e^{-2j \beta _i h_i})$ ($i=2, 3$). 
By combining \eqref{Ref_CO-a}, and \eqref{BCforH}, and after some straightforward mathematical manipulations, reflection coefficients can be derived as

\begin{subequations} \label{Ref-coefficients }
\begin{equation} \label{R00}
  {R_{00}} = 2\frac{{M_{00}^{22}{C_{22}} - M_{00}^{23}{C_{12}}}}{{{C_{22}}{C_{11}} - {C_{21}}{C_{12}}}} + 2\,\frac{{M_{00}^{33}\,{C_{11}} - M_{00}^{32}{C_{21}}}}{{{C_{22}}{C_{11}} - {C_{21}}{C_{12}}}}\, - 1  
\end{equation}

\begin{eqnarray} \label{Rm0}
    R_{m0}^{TM} = 2{\xi _{00}}({f_2}{S_2}A_{m0}^{ + ,2}\frac{{A_{00}^{ - ,2}{C_{22}} - A_{00}^{ - ,3}{C_{12}}}}{{{C_{22}}{C_{11}} - {C_{21}}{C_{12}}}} + \nonumber \\ {f_3}{S_3}A_{m0}^{ + ,3}\,\frac{{A_{00}^{ - ,3}\,{C_{11}} - A_{00}^{ - ,2}{C_{21}}}}{{{C_{22}}{C_{11}} - {C_{21}}{C_{12}}}})\,\,\, ; m \ne 0
\end{eqnarray}

\begin{eqnarray} \label{R0n}
    R_{0n}^{TE} = 2{\xi _{00}}({f_2}{S_2}A_{0n}^{ + ,2}\frac{{A_{00}^{ - ,2}{C_{22}} - A_{00}^{ - ,3}{C_{12}}}}{{{C_{22}}{C_{11}} - {C_{21}}{C_{12}}}} + \nonumber \\ {f_3}{S_3}A_{0n}^{ + ,3}\,\frac{{A_{00}^{ - ,3}\,{C_{11}} - A_{00}^{ - ,2}{C_{21}}}}{{{C_{22}}{C_{11}} - {C_{21}}{C_{12}}}})\,\,\, ; n \ne 0
\end{eqnarray}
\begin{eqnarray} \label{Rmn_tm}
  R_{mn}^{TM} =\frac{{2{\xi _{00}}k_{x,m}^2}}{{k_{x,m}^2 + k_{y,n}^2}}({f_2}{S_2}A_{mn}^{ + ,2}\,\frac{{A_{00}^{ - ,2}{C_{22}} - A_{00}^{ - ,3}{C_{12}}}}{{{C_{22}}{C_{11}} - {C_{21}}{C_{12}}}}+ \nonumber\\ f_3 S_3 A_{mn}^{+,3} \frac{{A_{00}^{ - ,3}\,{C_{11}} - A_{00}^{ - ,2}{C_{21}}}}{{{C_{22}}{C_{11}} - {C_{21}}{C_{12}}}});~~m,~n\ne0~~~~~~~
  \end{eqnarray}
\begin{equation} \label{Rmn_TE}
        R_{mn}^{TE} = \,R_{mn}^{TM}\frac{{k_{y,n}^2}}{{k_{x,m}^2}}  \,\,\,,\,\,\; m,n \ne 0 
\end{equation}
\end{subequations}
where
\begin{subequations} \label{C}
\begin{eqnarray} \label{C11}
   {C_{11}} = 0.5\,{S'_2}\, + M_{00}^{22} + M_{m0}^{22,TM} + M_{0n}^{22,TE} \nonumber \\ + M_{mn}^{22,TM} + M_{mn}^{22,TE}
\end{eqnarray}
\begin{equation} \label{C12}
    {C_{12}} = M_{00}^{32} + M_{m0}^{32,TM} + M_{0n}^{32,TE} + M_{mn}^{32,TM} + M_{mn}^{32,TE}
\end{equation}
\begin{equation} \label{C21}
    {C_{21}} = M_{00}^{23} + M_{m0}^{23,TM} + M_{0n}^{23,TE} + M_{mn}^{23,TM} + M_{mn}^{23,TE}
\end{equation}
\begin{eqnarray} \label{C22}
  {C_{22}} = 0.5\,{S'_3}\, + M_{00}^{33} + M_{m0}^{33,TM} + M_{0n}^{33,TE} \nonumber \\ + M_{mn}^{33,TM} + M_{mn}^{33,TE} 
\end{eqnarray}
\end{subequations}
and
\begin{subequations} \label{M}
\begin{equation} \label{M00}
    M_{00}^{ij} = {f_i}{S_i}{\xi _{00}}A_0^{ + ,i}\,A_0^{ - ,j}
\end{equation}

\begin{equation} \label{M_tm}
  M_{mn}^{ij,TM} = \frac{{k_{x,m}^2}}{{k_{x,m}^2 + k_{y,n}^2}}{f_i}{S_i}\xi _{mn}^{TM}A_{mn}^{ + ,i}\,A_{mn}^{ - ,j}\,\,\,\,\,;\,\,m,n \ne 0 
\end{equation}

\begin{equation} \label{M_te}
    M_{mn}^{ij,TE} = \frac{{k_{y,n}^2}}{{k_{x,m}^2 + k_{y,n}^2}}{f_i}{S_i}\xi _{mn}^{TE}A_{mn}^{ + ,i}\,A_{mn}^{ - ,j}\,\,\,\,;\,\,m,n \ne 0\,
\end{equation}
\end{subequations}
Finally, the diffraction efficiencies (the ratio of diffracted power to the incident wave) can be calculated by the following relation

\begin{subequations} \label{DE}
\begin{equation} \label{DE00}
D{E_{0,0}} = {\left| {{R_{00}}} \right|^2}\,
\end{equation}

\begin{eqnarray} \label{DE_tm}
DE_{m,n}^{TM} = {\left| {R_{mn}^{TM}} \right|^2}[1 + {(\frac{{{k_{y,n}}}}{{{k_{x,m}}}})^2} +\nonumber \\ {(\frac{{k_{x,m}^2 + k_{y,n}^2}}{{{k_{x,m}}{k_{z,mn}}}})^2}] ~ {\mathop{\rm Re}\nolimits} \{ \frac{{{k_{z,mn}}}}{{{k_{z,00}}}}\} \,\,\,\,\,\,;\,m \ne 0
\end{eqnarray}

\begin{equation} \label{DE_te}
DE_{m,n}^{TE} = {\left| {R_{mn}^{TE}} \right|^2}(1 + \frac{{k_{x,m}^2}}{{k_{y,n}^2}}){\mathop{\rm Re}\nolimits} \{ \frac{{{k_{z,mn}}}}{{{k_{z,00}}}}\} \,\,\,\,\,\,;\,n \ne 0
\end{equation}
\end{subequations}

\begin{figure} 
\centering\includegraphics[width=\linewidth]{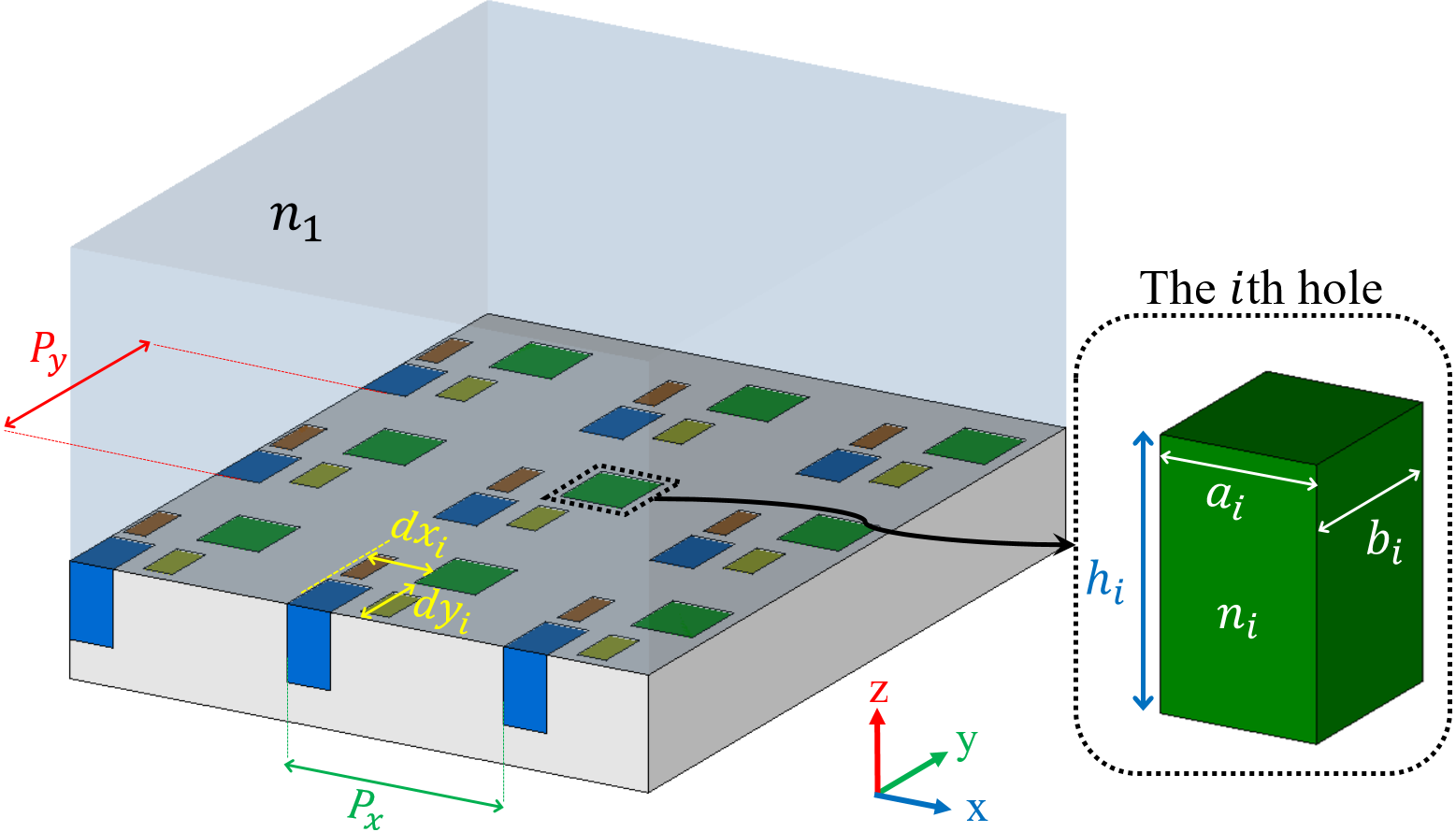}
\caption{Schematic illustration of the CMG consisting of multiple rectangular holes with an arbitrary arrangement. Inset: the $i$th holes.}
\label{CMG}
\end{figure}

These calculations can be generalized to the case of a 2D-CMG with more than two holes in each period. Fig.~\ref{CMG} depicts a 2D-CMG composed of $N$ (arbitrary number) holes per unit-cell with the lattice constant $P_x$ and $P_y$ along $x-$ and $y-$axes, respectively. We denote the corner of the $i$th hole by $d_{xi}$ and $d_{yi}$, its height by $h_i$ and width and length by $a_i$, and $b_i$, respectively (Fig.~\ref{CMG}). The $i$th hole is filled with a dielectric material with a refractive index of $n_i$. Similarly, the total electric and magnetic field must be expanded in all regions, and appropriate boundary conditions must be applied to derive the reflection coefficients of the diffracted orders. For brevity, the details of these calculations are not presented here.

\subsection{Numerical results}

\begin{figure}  
\centering\includegraphics[width=\linewidth]{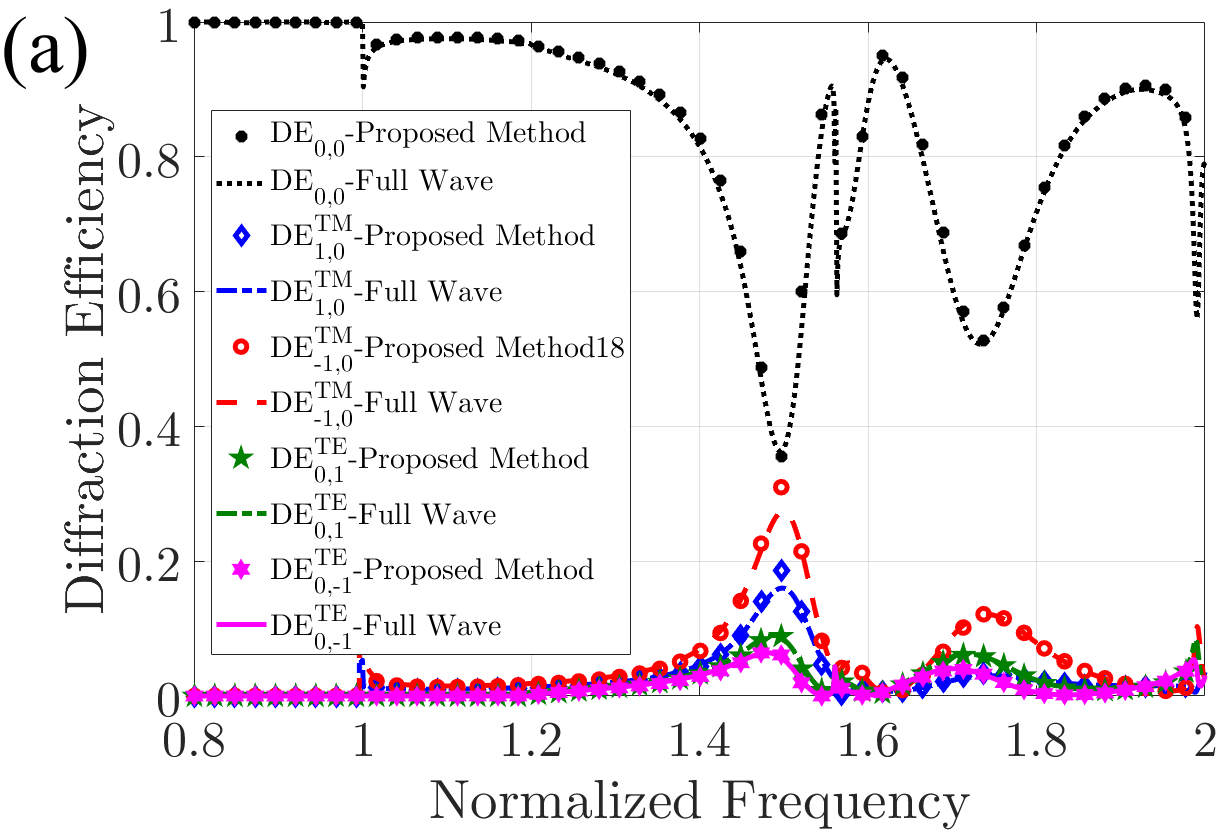}
\centering\includegraphics[width=\linewidth]{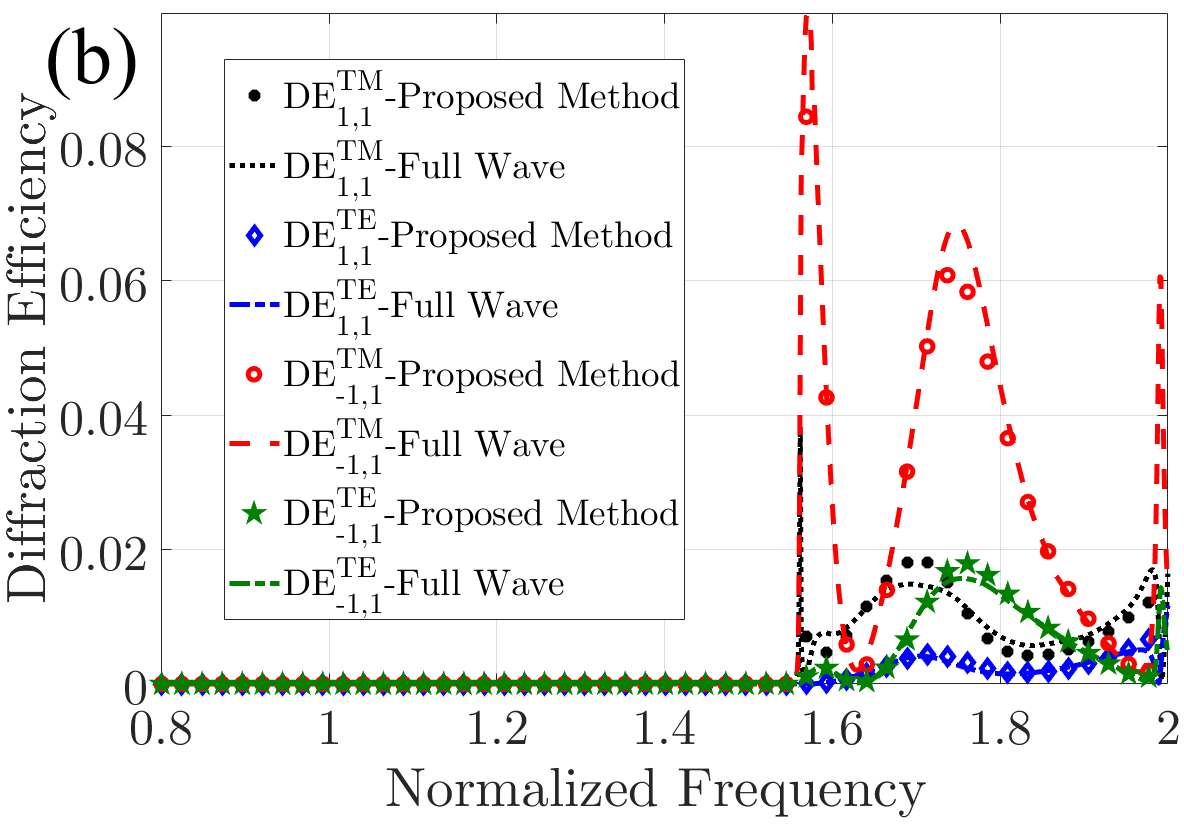}

\caption{ Comparing the results of the proposed method with full-wave simulations. The diffraction efficiency of (a) the five first (b) and the higher diffracted orders of CMG. The CMG parameters are assumed as $P_y=0.83 P_x$, $a_2=0.16 P_x$, $b_2=0.33 P_x$, $a_3=0.33 P_x$, $b_3=0.216 P_x$, $d_x=0.25 P_x$, $d_y=0$, $h_2=0.83 P_x$, $h_3=1.08 P_x$, and $n_1=n_2=n_3=1$.}
\label{NEx1}
\end{figure}

\begin{figure}  
\centering\includegraphics[width=\linewidth]{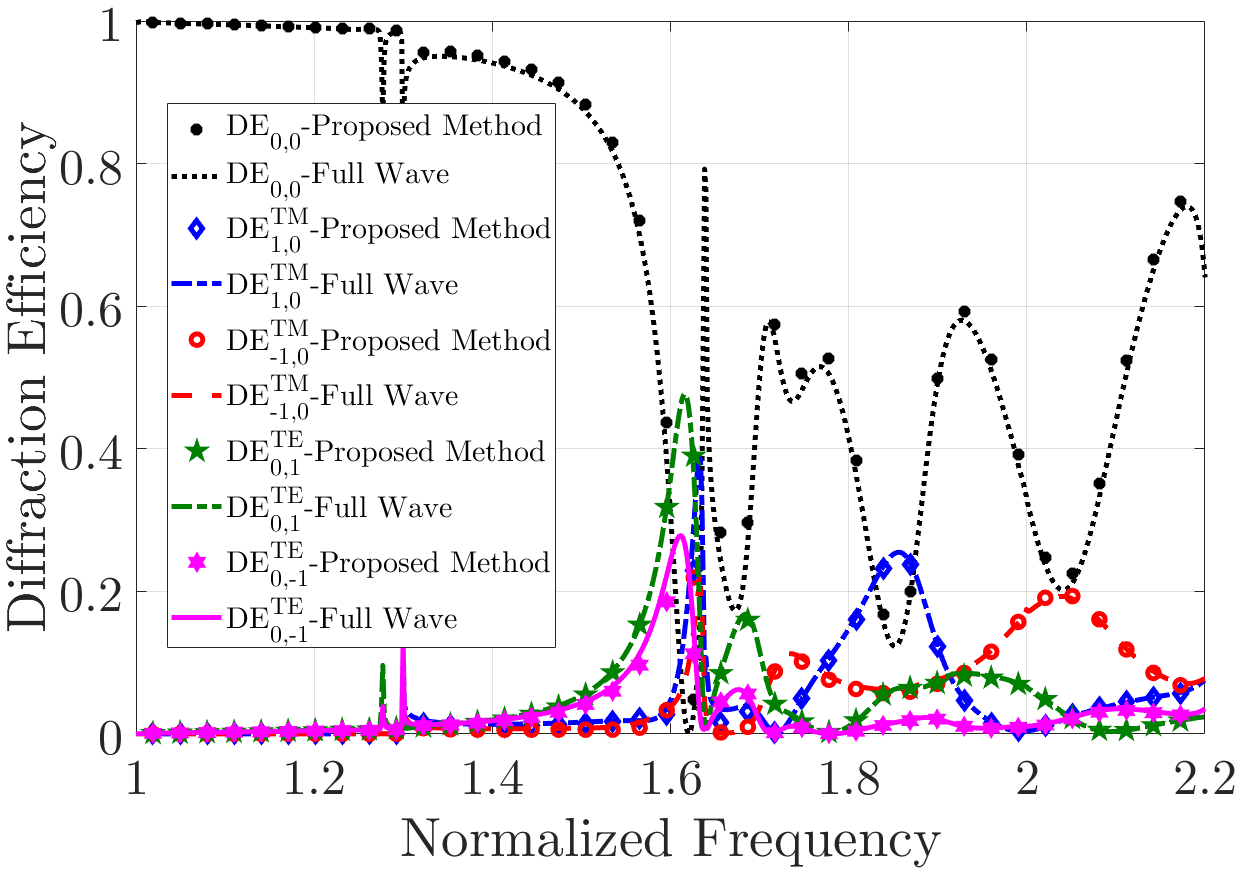}
\caption{The diffraction efficiency spectra of a 2D-CMG with four holes in each unit cell. The results are calculated by the proposed method and full-wave simulations. The geometrical parameters of CMG are mentioned in the text.}
\label{NEx2}
\end{figure}

Here, we present some numerical examples to verify the accuracy of the proposed method. As the first numerical example, in accordance with Fig.~\ref{CMG_twoH}, we set the parameter of the structure to $P_y=0.83 P_x$, $a_2=0.16 P_x$, $b_2=0.33 P_x$, $a_3=0.33 P_x$, $b_3=0.216 P_x$, $d_x=0.25 P_x$, $d_y=0$, $h_2=0.83 P_x$, $h_3=1.08 P_x$, and $n_1=n_2=n_3=1$. For the second example, consider a 2D-CMG with four holes in each period. The parameters of the structure are assumed as $P_x=0.77 P_y$, $a_2=b_2=a_3=a_4=0.154 P_y$, $b_3=0.277 P_y$, $b_4=b_5=0.3 P_y$, $a_5=0.185 P_y$, $h_2=0.615 P_y$, $h_3=0.77 P_y$, $h_4=0.69 P_y$, $h_5=0.92 P_y$, $d_{x2}=d_{y2}=d_{y3}=d_{x4}=0$, $d_{x3}=0.23 P_y$, $d_{y4}=0.3 P_y$, $d_{x5}=0.385 P_y$, $d_{y5}=0.46 P_y$, $n_1=n_3=n_4=n_5=1$, and $n_2=1.5$. The diffraction efficiencies of the diffracted orders versus the normalized frequency are displayed in Figs.~\ref{NEx1} and \ref{NEx2}. Here, we define the normalized frequency as $\omega_n=\max[P_x,P_y]/\lambda_0$, and $\lambda_0$ is the free space wavelength. A full-wave simulation is also carried out to validate the analytical method using the finite integration technique (FIT) in CST Microwave Studio 2019. In CST, periodic boundary conditions are applied in both $x$- and $y$-directions, while the perfectly matched layer (PML) boundary condition is applied in the $z$-direction. Evidently, the results of our proposed analytical method are in excellent agreement with those obtained by using the full-wave simulations.

\section{Applications and discussions}

\begin{figure} 
\centering\includegraphics[width=\linewidth]{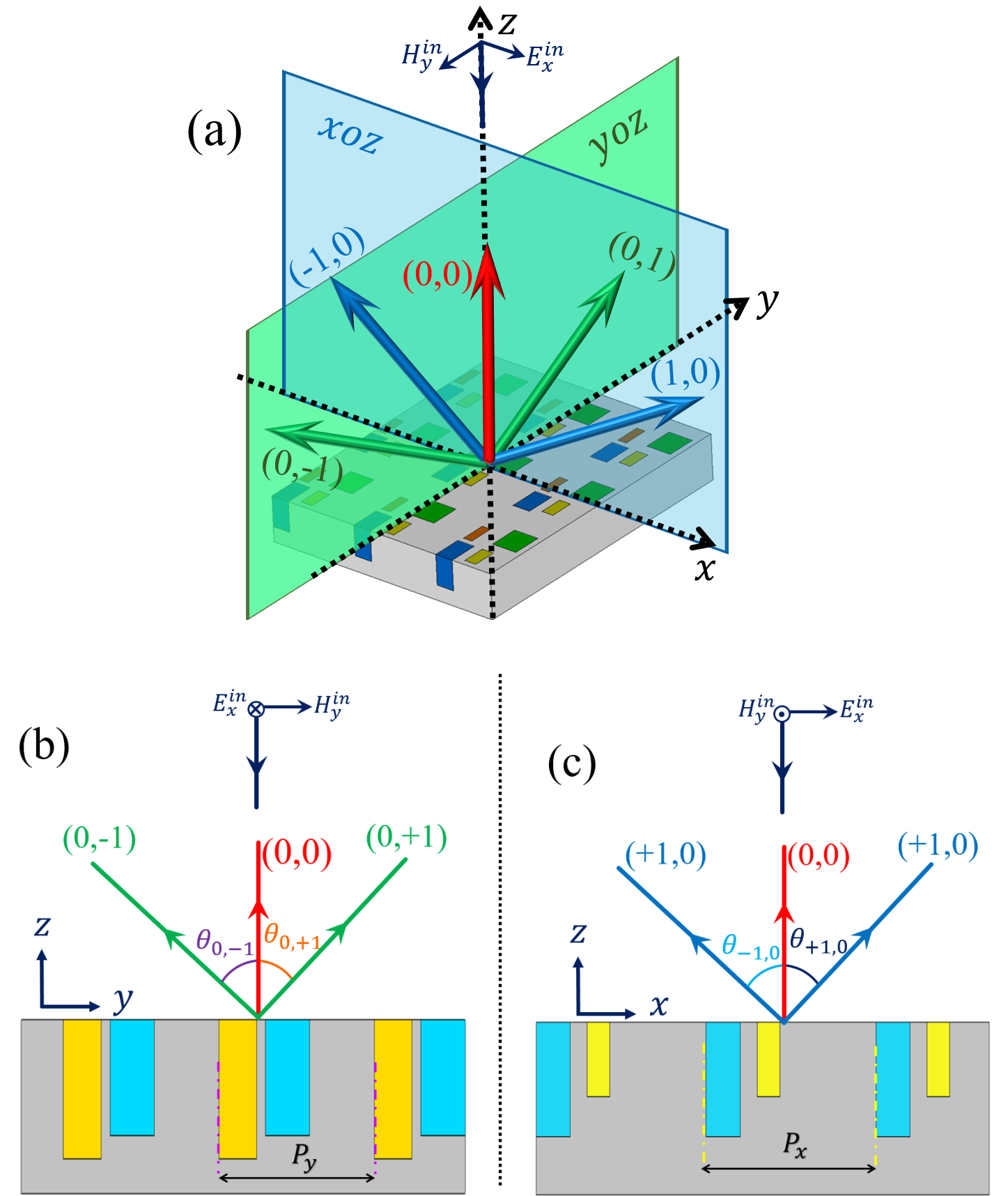}
\caption{ Schematic representation of 2D-metagratings with five propagating diffraction order. (a)3D view, (b) $y-z$ plane, and (c) $x-z$ plane.}
\label{2D-Metagratings}
\end{figure}

In this section, we design anomalous reflectors and beam splitters with near-unitary power efficiency using the proposed analytical method. As mentioned earlier, each term of the summations in \eqref{Ex_1}-\eqref{Hx_1} can be interpreted as a plane wave at elevation angle $\theta _{mn}$ and azimuth angle $\varphi_{mn}$ which differ from the angles of the incident wave ($\theta _{00}$ and $\varphi_{00}$ ), except for the specular mode ($00$ order). From \eqref{kxm1_kyn1-kz1}, $\theta _{mn}$ and $\varphi_{mn}$ of the diffracted wave can calculated by following equations
\begin{subequations} \label{phi-theta-mn}
\begin{equation} 
    \cos{\varphi_{mn}} \sin{\theta_{mn}}=k_{x,m}/k_0
\end{equation}
\begin{equation} 
     \sin{\varphi_{mn}} \sin{\theta_{mn}}=k_{y,m}/k_0
\end{equation}
\end{subequations}

The five first diffraction orders (i.e. $(0,0)$, $(\pm 1,0)$ and $(0, \pm 1))$ are depicted in Fig.~\ref{2D-Metagratings}, while the metagrating is illuminated by a normal TM plane wave. In showing how the structure works, we do not depict the higher diffraction orders in Fig.~\ref{2D-Metagratings} for simplification. In this case, $(\pm1,0)$ FB modes lie within the $x-z$ plane at angle $\theta_{\pm1,0}$ from the $z$-axis, and $(0,\pm1)$ FB modes lie within the $y-z$ plane at angle $\theta_{0,\pm1}$ from the $z$-axis as shown in Fig.~\ref{2D-Metagratings}. Moreover, the (0,0) mode overlaps at $z$-axis. Based on Equations \eqref{phi-theta-mn} and \eqref{kxm1_kyn1-kz1}, $\theta_{0,\pm1}$, and $\theta_{\pm1,0}$ can be expressed as
\begin{subequations} \label{theta}
\begin{equation}  \label{theta-10}
    |\theta_{\pm1,0}|= \sin^{-1}(\frac{\lambda_0}{Px})
\end{equation}
\begin{equation}  \label{theta-01}
     |\theta_{0,\pm1}|= \sin^{-1}(\frac{\lambda_0}{Py})
\end{equation}
\end{subequations}
and $\varphi_{\pm1,0}=0^\circ$ and $\varphi_{0,\pm1}=90^\circ$ for the normal incidence. Note that the azimuth angle of a higher diffraction order can take an arbitrary value due to periodicity and operating wavelength. Therefore, we have a multi-channel metagrating, with each channel having a certain elevation and azimuth angle proportional to the period of the structure and the wavelength. In the following, we aim to use a 2D-CMG to manipulate the power distribution between these channels of this metagrating, achieve the desired diffraction pattern, and propose various applications accordingly.

\subsection{Perfect out-of-plane anomalous reflection}
\begin{figure*} 
\centering\includegraphics[width=16.5cm]{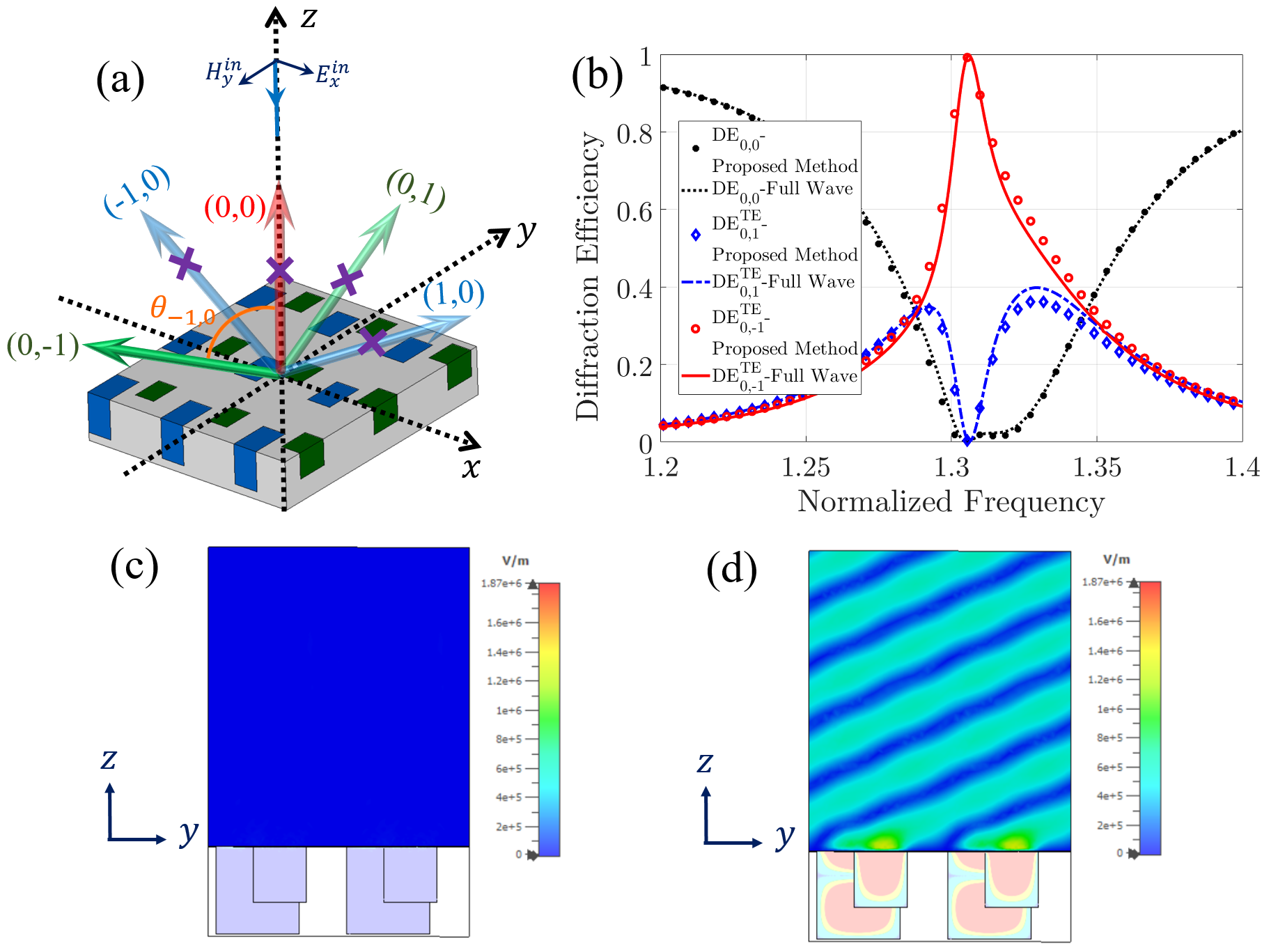}
\caption{(a) Schematic representation of the proposed out-of-plane reflector. (b) Diffraction efficiency of the designed perfect out-of-plane reflector. Distribution of the absolute value of (c) $E_y$, and (d) $E_x$ in $y-z$ plane for $\omega_n= 1.305$ (design frequency) . The 2D-CMG parameters are designed as $P_y=1.3\lambda_0$, $P_x=0.59 P_y$, $a_2=0.083 P_y$, $a_3=0.073 P_y$, $b_2=0.637 P_y$, $b_3=0.4 P_y$, $h_2=0.669 P_y$, $h_3=0.427 P_y$, $d_x=0.46 P_y$, and $d_y=0.288 P_y$. Full-wave simulation results and E-field distribution are obtained by CST Microwave Studio.}
\label{AR}
\end{figure*}

In this subsection, we design a perfect out-of-plane anomalous reflector using the proposed analytical method. The problem geometry is depicted in Fig.~\ref{AR}(a). Our goal is to couple the normal incident TM plane wave, to a TE plane wave in the $y-z$ plane with angle $\theta_{0,-1}$ from the $z$-axis. Note that, $(0,\pm1)$ FB mode are propagation along a direction in the $y-z$ plane which dose not lie in the plane of incidence ($x-z$ plane). Hence, if the power of incident wave transfer to $(0,\pm1)$ FB mode, out-of-plane reflection can be realized.

To simplify the design process, we assume that only the $(0,0)$ and $(0,\pm1)$ are propagating while higher-order diffracted modes are evanescent, which is achieved by choosing $P_x$,and $P_y$ from the range of $[0,\lambda_0]$ and $[\lambda_0, 2 \lambda_0]$, respectively, and satisfying the following condition : $k_{x,1}^2+k_{y,1}^2 < k_0^2$. By eliminating the $DE_{0,0}$ and $DE_{0,1}^{TE}$, we achieve unitary efficiency for the $(0,-1)$ mode since the higher-order modes are evanescent and the entire structure is lossless. Based on \cite{popov2018controlling,popov2019constructing}, for perfect elimination of $N$ FB modes, $N$ meta-atoms are needed for the structure of 1D metagrating. According to \cite{rabinovich2020dual}, this principle is verified for 2D metagratings. Here, each hole is considered as a meta-atom; therefore, to suppress two FB modes, we use a 2D-CMG with two holes per period.

For a $-50^\circ$ deflection angle, based on \eqref{theta-01}, the periodicity of the structure along the $y$-axis must be chosen as $1.3 \lambda_0$. To further simplify the fabrication process, we assume that all holes are filled with air ($n_i=1$). To extract the other parameters, we utilize the genetic algorithm (GA) to minimize the $DE_{0,0}$ and $DE_{0,1}^{TE}$ of the structure. Using the proposed method, we define the cost function as $DE_{0,0}+$ $DE_{0,1}^{TE}+$ $1/DE_{0,-1}^{TE}$ in the desired frequency. The optimized parameters of the structure are extracted as $P_x=0.59 P_y$, $a_2=0.083 P_y$, $a_3=0.073 P_y$, $b_2=0.637 P_y$, $b_3=0.4 P_y$, $h_2=0.669 P_y$, $h_3=0.427 P_y$, $d_x=0.46 P_y$, and $d_y=0.288 P_y$. The diffraction efficiencies of the optimized metagratings are plotted in Fig.~\ref{AR}(b), depicting an excellent agreement between the results of full-wave simulation and those predicted by our analytical approach. It can be seen in Fig.~\ref{AR}(b) that almost all the power of the incident wave ($99.9\%$) is transferred to the $(0,-1)$ order in the desired frequency $\omega_n =1.305$. This efficiency is a remarkable achievement compared with previously reported anomalous reflectors \cite{zhou2020polarization,chen2018polarization,inampudi2018neural,rabinovich2020dual,aieta2012out}. The magnitude of the electric field distributions is also depicted in Figs.~\ref{AR}(c) and \ref{AR}(d). Based on the electric field distributions, the designed metagratings transfer a normal incident TM plane wave to an oblique TE plane wave (with an angle of $\theta_{0,-1}=-50^\circ$) in the $y-z$ plane.

\begin{table*} 
\caption{Optimum parameters for the perfect out-of-plane reflectors using 2D-CMG. For simplifying the fabrication process, the holes are filled with air.}
\begin{center}
\begin{tabular}{p{1.25cm} p{1.25cm} p{1.25cm} p{1.25cm} p{1.25cm} p{1.25cm} p{1.25cm} p{1.25cm} p{1.25cm} p{1.25cm} p{1.25cm}} 
 \hline
 $\theta_{0,-1}$ & $P_x/P_y$ &$a_2/P_y$ & $b_2/P_y$ & $a_3/P_y$ & $b_3/P_y$ & $h_2/P_y$ & $h_3/P_y$ & $d_{x}/P_y$ & $d_{y}/P_y$ & PE(\%) \\  \hline
  $-75^\circ$ & 0.8 & 0.2 & 0.48 & 0.14 & 0.53 & 1 & 0.39 & 0.2 & 0.1& 98.6 \\ \hline
   $-65^\circ$ & 0.75 & 0.22 & 0.48 & 0.04 & 0.46 & 0.5 & 0.9 & 0.25 & 0.16& 99.6 \\ \hline
 $-55^\circ$ & 0.32 & 0.23 & 0.7 & 0.04 & 0.44 & 0.75 & 0.47 & 0.27 & 0.26 & 99.9 \\ \hline
  $-45^\circ$ & 0.35 & 0.08 & 0.4 & 0.1 & 0.38 & 0.41 & 0.4 & 0.24 & 0.27 & 99.9 \\ \hline
$-35^\circ$ & 0.51 & 0.44 &  0.32 & 0.16 & 0.28 & 0.46 & 0.51 & 0.05& 0.35 & 99.2 \\ \hline
\end{tabular}
\end{center}
\label{AR-table}
\end{table*}

Moreover, the 2D-CMG can be used for designing out-of-plane reflectors with different design angles. Similarly, we repeat the design process for extracting parameters of the anomalous reflector with $\theta_{0,-1}$ in the range of $-35^\circ$ to $-75^\circ$. The optimized structure parameters and the power efficiency (PE) of the designed metagratings are listed in Table~\ref{AR-table}. In all of the designed anomalous reflectors, we can achieve near-unitary efficiency. It should be noted that according to \eqref{kxm}, the anomalous reflection occurs in the normalized frequency $\omega_n=P_y/\lambda_0$. Note that the 2D-CMG can also be used to realize the in-plane anomalous reflectors (coupling the incident power to ($\pm1,0), (\pm2,0), ...$ FB modes ). Nevertheless, there are some structures with less complication, such as 1D-CMG, that can realize this phenomenon \cite{rahmanzadeh2020perfect}. For brevity, we do not present the results of this application here.

\subsection{Five-channel beam splitters}

\begin{figure*} 
\centering\includegraphics[width=18cm]{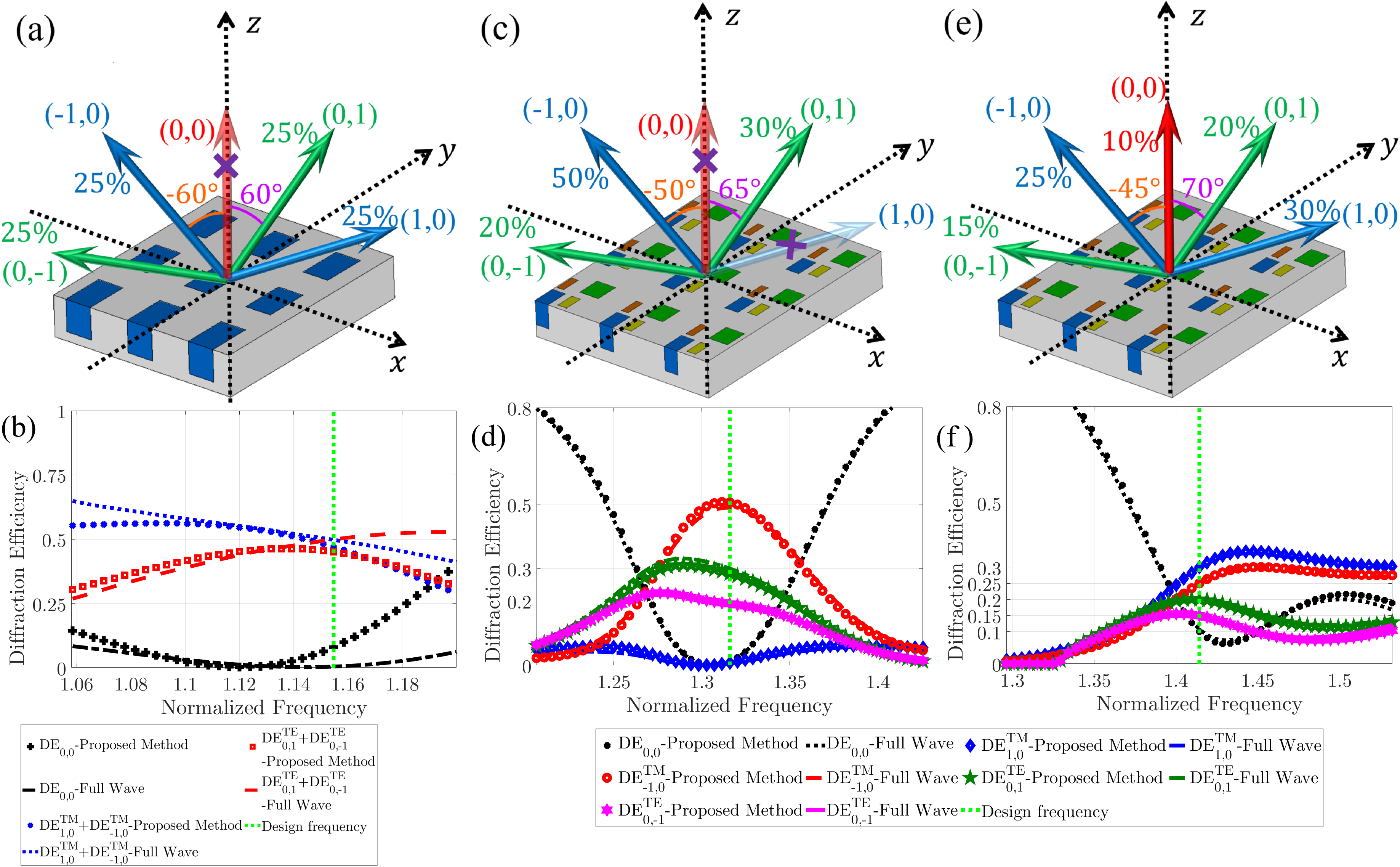}
\caption{ (a), (c), and (e) The schematics of the proposed five-channel beam splitters. (b), (d), and (f) The diffraction efficiencies of the designed beam splitters. The incident power can be arbitrarily distributed between diffraction orders by the 2D-metagratings. The optimum parameters of the structure are mentioned in the text and also listed in Table~\ref{BS_2}, and Table~\ref{BS_3}. Based on the given design angle, beam splitting occurs in the normalized frequency $\omega_n=1.115$, $\omega_n=1.316$, $\omega_n=1.414$ for the first, second, and third beam splitters, respectively.}
\label{BeamSplitter}
\end{figure*}

Herein, we design several five-channel beam splitters with an arbitrary power distribution based on the concept of metagratings using analytical expressions derived in the previous section. The new problem geometry is illustrated in Fig.~\ref{BeamSplitter}. To simplify the design process, we assume that only the first five diffraction orders ($(0,0)$, $(0,\pm1)$, and $(\pm1,0)$) are propagating, and higher orders do not carry any power in region 1. Consequently, we restrict the periodicities of the structure and the operating frequency to the range that satisfies these conditions: $k_{x,1}^2+k_{y,1}^2 < k_0^2$, $k_{y,2}^2 < k_0^2$, and $k_{x,2}^2 < k_0^2$. By distributing the incident power between these diffraction orders, a five-channel beam splitter can be realized. Here, we note again that two channels (orders) of the proposed beam splitter lie in the $x-z$ plane$(\pm1,0$), two of them lie in the $y-z$ plane $(0,\pm1)$, oriented along the angle $\theta_{\pm1,0}$ and $\theta_{0,\pm1}$, respectively, from the z-axis, and these angles can be controlled by changing the periodicities. Furthermore, the final channel $(0,0)$ overlaps on the $z$-axis. To attain desired power distribution and complete control over diffraction patterns, four meta-atoms (holes) per period can provide sufficient degrees of freedom, based on what was outlined in\cite{popov2018controlling,popov2019constructing} and due to the passivity condition (Note that the proposed structure is lossless).

As proof of the concept, we designed three devices for five-channel beam splitting using the proposed analytical method. All holes of the designed beam splitters in this subsection are filled with air for more simplicity in the fabrication process ($n_i=1$). The first beam splitter suppresses the $(0,0)$ FB mode and divides the incident power uniformly into four directions: two of which lie in the $x-z$ plane, and the others lie in the $y-z$ plane. All directions have an angle of $60^\circ$ to the $z$-axis ($|\theta_{\pm1,0|}$=$|\theta_{0,\pm1}|$=$60^\circ$). Since this beam splitter has a symmetric diffraction pattern ($DE_{0,1}^{TE}=DE_{0,-1}^{TE}$ and $DE_{1,0}^{TM}=DE_{-1,0}^{TM}$), for a simpler design, we can use a metallic grating with one rectangular hole in each unit cell as shown in Fig.~\ref{BeamSplitter}(a). According to the given angle and \eqref{theta}, periodicities must be chosen as $P_x=$ $P_y=1.155 \lambda_0$. After running an optimization using the proposed method (here the cost function defined as $|DE_{0,1}^{TE}-0.25|^2$ $+|DE_{1,0}^{TM}-0.25|^2$ in the normalized frequency $\omega_n=1.115$), the other parameters of the structure (i.e., the width, the length, and the height of the rectangular hole) are extracted as $a_2=0.65 P_x$, $b_2=0.479 P_x$, and $h_2=0.564 P_x$. The diffraction efficiencies of the optimized structure are plotted in Fig.~\ref{BeamSplitter}(b). Evidently, power uniformly is transferred to $(0,\pm1)$, and $(\pm1,0)$ in the
desired frequency $(\omega_n=1.115)$. The relative distribution error (defined here as relative deviation from the desired power distribution) is less than $1\%$, and the total diffraction efficiencies of orders are more than $99.9\%$, which is significantly improved in terms of both the power efficiency and relative distribution error compared with the previously published beam splitters \cite{lv2020all,pang2019alternative,gong2020polarization,zhou2021polarization,zhang2018nanoscale}.

\begin{table} 
\caption{Optimum parameters for the second designed beam splitter using 2D-CMG. The holes are filled with air to simplify the fabrication process.}
\begin{center}
\begin{tabular}{c c c c c c c} 
 \hline
 {$a_2/P_x$} & {$b_2/P_x$} & {$h_2/P_x$} & {$d_{x2}/P_x$} & {$d_{y2}/P_x$} & {$a_3/P_x$} & {$b_{3}/P_x$} \\ \hline  0.09 & 0.25 & 0.15 & 0 & 0 & 0.35 & 0.31 \\ \hline {$h_{3}/P_x$} & {$d_{x3}/P_x$} & {$d_{y3}/P_x$} & {$a_4/P_x$} & {$b_4/P_x$} & {$h_4/P_x$} & {$d_{x4}/P_x$} \\ \hline  0.77 & 0.62 & 0 & 0.13 & 0.4 & 0.73 & 0 \\ \hline {$d_{y4}/P_x$} & {$a_5/P_x$} & {$b_{5}/P_x$} & {$h_{5}/P_x$} & {$d_{x5}/P_x$} & {$d_{y5}/P_x$} \\ \hline 0.41 & 0.45 & 0.4 & 0.74 & 0.25 &  0.34 \\ \hline
\end{tabular}
\end{center}
\label{BS_2}
\end{table}

\begin{table} 
\caption{Optimum parameters for the third designed beam splitter using 2D-CMG. The holes are filled with air to simplify the fabrication process.}
\begin{center}
\begin{tabular}{c c c c c c c} 
 \hline
 {$a_2/P_x$} & {$b_2/P_x$} & {$h_2/P_x$} & {$d_{x2}/P_x$} & {$d_{y2}/P_x$} & {$a_3/P_x$} & {$b_{3}/P_x$} \\ \hline  0.31 & 0.1 & 0.37 & 0 & 0 & 0.26 & 0.09 \\ \hline {$h_{3}/P_x$} & {$d_{x3}/P_x$} & {$d_{y3}/P_x$} & {$a_4/P_x$} & {$b_4/P_x$} & {$h_4/P_x$} & {$d_{x4}/P_x$} \\ \hline  0.37 & 0.55 & 0 & 0.16 & 0.37 & 0.53 & 0 \\ \hline {$d_{y4}/P_x$} & {$a_5/P_x$} & {$b_{5}/P_x$} & {$h_{5}/P_x$} & {$d_{x5}/P_x$} & {$d_{y5}/P_x$} \\ \hline 0.35 & 0.37 & 0.33 & 0.43 & 0.4 &  0.33 \\ \hline
\end{tabular}
\end{center}
\label{BS_3}
\end{table}

In the following, we design two beam splitters with an asymmetric diffraction pattern, unlike the first beam splitter. Therefore, in these cases, we use a 2D-CMG that has four holes in a period. A schematic representation of the second beam splitter is depicted in Fig.~\ref{BeamSplitter}(c). This beam splitter eliminates $(0,0)$ and $(1,0)$ FB modes while transferring $50\%$ of the incident power to $(-1,0)$ order, $30\%$ to $(0,+1)$ order, and $20\%$ to $(0,-1)$. In this case, $(\pm1,0)$ orders are oriented along directions with $\theta_{\pm1,0}=\pm 50^\circ$ to the $z$-axis and $(0,\pm1)$ orders along directions with $\theta_{0,\pm1}=\pm 65^\circ$ to the $z$-axis. Hence, $P_x$ and $P_y$ must be chosen as $1.3 \lambda_0$ and $1.1 \lambda_0$, respectively, according to \eqref{theta}. The third designed metagrating reflect $10\%$ of the incident power to the specular mode and transfer $30\%$ and $25\%$ of the incident power to the channels lying in the $x-z$ plane with angle $\theta_{\pm1,0}=$ $\pm45^\circ$ to the $z$-axis. The rest of the incident power goes to the $y-z$ plane with a splitting ratio of $3:4$ and angle $\theta_{0,\pm1}=$ $\pm70^\circ$ as shown in Fig.~\ref{BeamSplitter}(e). To design such a beam splitter, periodicity along the $x$- and $y$-axes must be chosen as $1.41 \lambda_0$ and $1.064 \lambda_0$, respectively, according to \eqref{theta}.
Again, we utilize GA to extract other parameters of the second and third beam splitters using the proposed analytical method (Tables~\ref{BS_2} and \ref{BS_3}). The used cost function for designing these beam splitters were $|DE_{-1,0}^{TM}-0.50|^2$ $+|DE_{0,1}^{TE}-0.3|^2$ $+|DE_{0,-1}^{TE}-0.2|^2$ and $|DE_{0,0}-0.1|$ $+|DE_{1,0}^{TM}-0.3|$ $+|DE_{-1,0}^{TM}-0.25|^2$ $+|DE_{0,1}^{TE}-0.20|^2$ $+|DE_{0,-1}^{TE}-0.15|^2$, respectively for the second and third cases. The diffraction efficiencies of the designed metagratings are plotted in Figs.~\ref{BeamSplitter}(d) and \ref{BeamSplitter}(f), depicting an excellent agreement between the results of the full-wave simulation and our analytical method. The findings demonstrate a near-unitary total efficiency $(99.9\%)$ and a relative distribution error of $<1\%$, which is a remarkable achievement compared with previously reported grating- and metasurface-based beam splitters \cite{popov2019designing,popov2019constructing,popov2018controlling,zhou2021polarization,zhang2018metasurface,chen2021all,wang2021multi,tian2020nanoscale,li2020ultra}. Therefore, a five-channel beam splitter with arbitrary power distribution and a near-unitary efficiency can be realized using the proposed method and based on the concept of metagrating.

\begin{figure*} 
\centering\includegraphics[width=16.5cm]{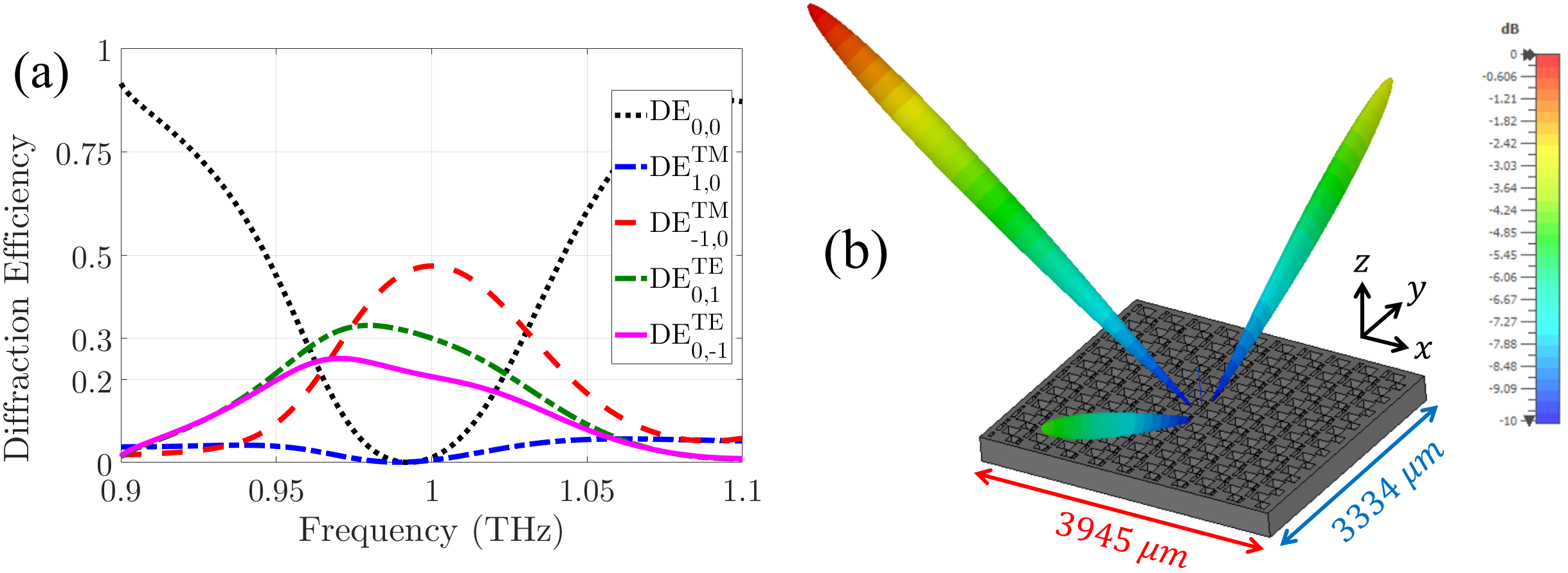}
\caption{ (a) Diffraction efficiency of the proposed metagrating that splits the incident wave with the desired ratio. (b) The far-field scattering patterns (normalized) of the designed beam splitter at 1 THz. In this case, unlike the previous examples, we use a lossy metal for metagrating design. ($\sigma =5.96$ $\times 10^7 $). CST Microwave Studio 2019 was utilized to extract the performance of this beam splitter.}
\label{RCS}
\end{figure*}

Next, we consider the effects of the metal ohmic losses on our proposed devices. As noted earlier, we assume that rectangular holes in 2D-CMG are carved on a PEC slab (not a real metallic slab). This approximation is valid for microwave, millimeter-wave, and low THz regimes. Hence, the designed metagratings can be used in a wide range of frequencies. To investigate this effect, we replace PEC with a lossy metal (copper with a conductivity of $\sigma =5.96$ $\times 10^7 $) in the second designed beam splitter. We consider operating frequency to be 1THz, and other parameters of the structure can obtain from Table~\ref{BS_2}. We perform a full-wave simulation to plot diffraction efficiencies versus frequency (Fig.~\ref{RCS}(a)). The results do not strictly change, and even in this case, the relative distribution error and the total power efficiency are $2.15\%$, and $99.9\%$, respectively, which are better than the previously reported beam splitters. The performance of other designed metagratings with lossy metal is similar, and their results are not presented here for brevity.

Finally, we investigate the diffraction pattern of the designed metagrating when truncated to a finite size. Again, we only investigate the second beam splitter made with the lossy metal. To extract the scattering patterns of the 2D-CMG with a finite size, we perform a 3D simulation using CST Microwave Studio 2019. The physical size of CMG is approximately $3.95 mm$ in the $x$-direction and $3.33 mm$ in the $y$-direction ($10 \times 10$ unit cell). The truncated metagrating is under a normal TM-plane wave and the far-field patterns are depicted in Fig.~\ref{RCS}(b) at 1THz. The results show that almost no power is transferred to the directions with angles $\theta=$ $\varphi=0^\circ$ ((00) FB mode) and angles $\theta=50^\circ$, $\varphi=0^\circ$ ((10) FB mode) as we expected from Fig.~\ref{RCS}(a). Also, it can be observed that in the operating frequency, the incident power is split into three desired directions with a predesigned ratio.

\section{Conclusion}

Herein, a 2D-CMG was proposed for manipulating in-plane and out-of-plane EM waves based on the concept of metagratings. An analytical method was introduced for diffraction analysis of 2D-CMGs and verified through some numerical examples, indicating excellent agreement with full-wave simulation results. Closed-form and analytical expressions were also presented for the diffraction efficiency of the diffracted orders. By using the proposed method and without needing a single simulation in the full-wave software, we designed out-of-plane reflectors and five-channel beam splitters. The proposed reflectors transferred a normal TM plane wave to an oblique TE plane wave in the $y-z$ plane with angles above $-30^\circ$ to the $z$-axis with unitary power efficiency. The designed beam splitter distributed the incident power to five directions with an arbitrary ratio. The total power efficiency of the proposed beam splitters was above $99.9\%$ and their relative distribution error was less than $1\%$. This proposed method can pave the way for the analytical design of 2D metagratings with various potential applications for microwave and terahertz wavefront manipulation.

\bibliography{main}

\end{document}